**Mineral False Positives in the Search for Exoplanet Surface Biosignatures**

Parkinson, M.B., Kaltenegger, L., Biller, B., Lach, G., & McMahon, S.

## 0. Abstract

In the search for life in the cosmos, biopigments on exoplanet surfaces are a critical target. Such pigments have been detected in Earth's spectrum (by the Galileo spacecraft and in Earthshine) via the "vegetation" or "photosynthesis red edge" (VRE or PRE), a sharp, step-like increase in reflectance with increasing wavelength at ~700 nm. Future space telescopes like the Habitable Worlds Observatory (HWO) are designed to obtain disk-integrated spectra of Earth-like exoplanets in the visible-to-near-infrared to identify such features. However, there has been no systematic analysis of the occurrence of similar reflectance edges among minerals of non-biological origin. Here, we use existing databases of mineral reflectance spectra to explore the risk that minerals may present false positives in the search for biopigments on exoplanets. We find that several sulfide and tectosilicate minerals, as well as the prebiotically important cyanide salt, potassium ferrocyanide, have PRE-like features. We characterize these features in order to assess how they may be distinguished from biopigments. We conclude that the future evaluation of the biogenicity of PRE-like features in exoplanet reflectance spectra can be informed by the atmospheric context, but may require an assessment of the prior probability of non-biological and biological hypotheses about the surface materials of exoplanets.

## 1. Introduction

The search for biosignatures on extrasolar planets can encompass both the transmission spectra of their atmospheres and the spectral signatures of surface materials observable in reflected light (e.g., DesMarais et al., 2002; Kaltenegger et al., 2017; Schwieterman et al., 2018). Because Earth-sized planets in the habitable zone (e.g., Kasting, 1993; Kopparapu et al., 2014; Ramirez, 2018; Bohl et al., 2025) are characterized by low contrast ratios (brightness of the planet compared to brightness of the star), especially in the visible and near infrared, direct imaging of their disks is currently precluded by the limitations of existing telescopes (e.g., Currie et al., 2023).

Recognizing the compelling scientific potential for directly detecting extraterrestrial biomass through reflected light, the most recent National Academies' Decadal Survey for astronomy and astrophysics (National Academies of Sciences & Medicine, 2021) prioritized the development of a large infrared/optical/ultraviolet (IR/O/UV) space telescope. The survey emphasized that this mission, as a top priority within a Great Observatories Mission and Technology Maturation Program, should be designed to directly search for signatures of life on approximately 25 habitable-zone planets while also transforming general astrophysics. This recommendation led to NASA's proposal for the flagship Habitable Worlds Observatory (HWO), scheduled for launch in the early 2040s (e.g., Dressing et al., 2024). HWO aims to balance the ambitions of previous mission concepts, notably HabEX (Habitable Exoplanet

Observatory) (Gaudi et al., 2020), and LUVOIR (Large UV/Optical/IR Surveyor) (LUVOIR Team, 2019), which proposed to image ExoEarths in the visible and UV range using an internal coronagraph and an external starshade (occulter). HWO is planned to have a roughly 6-meter aperture and a coronagraph that is designed to enable imaging of planets 10 billion times fainter than their star (e.g., Dressing et al., 2024) over the near-IR/O/UV spectral range. This will be crucial for studying the atmospheres and surface reflectance properties of Earth-like planets in their habitable zones. Perhaps surprisingly, existing models suggest that such a telescope should be capable of detecting biopigments on a planet's surface via its contribution to the reflectance spectrum (e.g., Seager et al., 2005; O'Malley-James & Kaltenegger, 2017, 2018; Borges et al., 2024; Metz et al., 2024; Pham & Kaltenegger, 2021, 2022).

The paradigmatic example of a planetary reflectance biosignature is the "vegetation red edge" (VRE), also more generally known as the "photosynthetic red edge" (PRE; O'Malley-James & Kaltenegger, 2019) because photosynthetic pigments also occur in non-plants. This sharp, step-like increase in reflectance from wavelengths below ~700 nm to wavelengths above ~700 nm is characteristic of most photosynthetic organisms, which absorb predominantly in the "photosynthetically active" visible range (e.g., Coelho et al., 2022). This feature can be resolved in Earth's global spectrum by spacecraft instruments, is recoverable from Earthshine on the moon (e.g., Sagan et al., 1993; Seager et al., 2005; Arnold et al., 2009), and will targeted on extrasolar Earth-like planets (Seager et al., 2005; Fujii et al., 2010; Kaltenegger, 2017; Fuji et al., 2018; Schwieterman et al., 2018; Coelho et al., 2024). Some forms of photosynthetic biomass display analogous step-like reflectance "edges" at other wavelengths (e.g., Hegde et al., 2015; Coelho et al., 2022, 2024). The red edge and these other edges arise physically from a combination of strong absorption by pigments and strong scattering by cells and tissues (e.g., O'Malley-James & Kaltenegger, 2018 and references therein).

Historically, biosignature detection in astrobiology has often been frustrated by "false positives" resulting from the misinterpretation of non-living materials and processes (e.g., McMahon & Cosmidis, 2021; Vickers et al., 2023). To avoid this problem in the search for signals of life on exoplanets, it is essential to consider the abiotic baseline (e.g., Constantinou et al., 2025). In the context of surface reflectance, this requires an assessment of whether and how non-biological materials on planetary surfaces may mimic the VRE/PRE and other biomass-like spectral "edges" (e.g., Hedge et al., 2015; Coelho et al., 2022, 2024; Borges et al., 2024; Rodriguez et al., 2024).

Remote sensing specialists have long been aware of the potential for minerals to produce VRE/PRE-like features in reflectance spectra acquired from Earth (e.g., Clark, 1999). Semiconductor materials typically exhibit spectral reflectance edges in the visible to near-infrared wavelengths, corresponding to the "band gap", i.e., the energy above which incident photons are absorbed more readily (i.e., reflected less) because they can excite valence electrons up to the conduction band (Seager et al., 2005). Sulfur and several sulfide compounds are known to be semiconductors. Seager et al. (2005) pointed out that both crystalline sulfur (band gap ~450 nm) and cinnabar (HgS; 600 nm) may therefore present red-edge-like false positives, but noted that "the wavelength of Earth's vegetation red edge does not correspond to that of any known mineral". More recently, Borges et al. (2024) drew attention to steep spectral slopes in iron oxide and hydroxide minerals as possible sources of VRE/PRE-like false positives. However, to our knowledge there has been no systematic search for VRE/PRE-like features in the reflectance spectra of minerals. This paper presents the results of such a search together with simulated visible-IR reflectance spectra from exoplanets with high surface coverage of such minerals.

## 2. Methods

### *2.1 Spectral libraries*

We obtained spectra from the U.S. Geological Survey (USGS) Spectral Library (Kokaly et al., 2017) and the NASA ECOSTRESS database (Meerdink et al., 2019). From the USGS database, we sourced spectra from 290 artificial materials and 1,276 minerals, which includes arsenates (1 spectrum), borates (6), carbonates (17), halides (2), hydroxides (10), oxides (13), phosphates (5), silicates (121), sulfides (8), and sulfates (20). Additionally, 1,040 vegetation spectra and 1,609 mineral spectra were obtained from the NASA ECOSTRESS database, which includes arsenates (18 spectra), borates (62), carbonates (158), halides (48), hydroxides (19), oxides (106), phosphates (36), silicates (897), sulfides (99), and sulfates (137). All the spectra in these databases were originally measured using benchtop spectrometers under controlled conditions, providing absolute reflectance (0–100%). Multiple spectra for the same mineral were often available capturing variations due to grain size or the choice of instrumentation and measurement conditions. We determined the spectral resolution of the USGS and ECOSTRESS datasets by measuring the difference between consecutive wavelength measurements. The mean resolution of all USGS spectra data is 9.88 nm and the mean resolution of all ECOSTRESS data is 0.822 nm.

To examine and visually compare spectra, we used the open-source Quasar software (Toplak et al., 2021). For visualization only (e.g., in **Figure 1**), we applied a Gaussian smoothing filter with a standard deviation (SD) of 3 to the data to minimize high-frequency noise. We manually examined all mineral spectral data to assess quality and to identify any features within the 500–2,500 nm range relevant to the study (grain sizes of minerals are given in **Table 1**: coarse (>125μm), medium (45–125μm), and fine (<45μm)).

To provide a coherent red edge comparison throughout the paper, we downloaded the green leaf spectrum from the RELAB facility, Brown University's online catalog (http://www.planetary.brown.edu/relab/; Green Maple Leaf C1GRLF). This is the same green leaf spectrum the Planetary Spectrum Generator uses (section 2.3).

### *2.2 Spectral treatment and edge detection*

To search for VRE/PRE-like features, we trimmed the spectra to the 600–800 nm region and wrote an algorithm in Python to search for strong increases in reflectance (with wavelength) between pairs of wavelength values ($\lambda_{end} > \lambda_{start}$), i.e., 'edges'. In order to find edges that mimic the VRE/PRE, the wavelength interval between $\lambda_{start}$ and $\lambda_{end}$ was set to a maximum of 80 nm (excluding gentle ramps in favor of 'edges') and a minimum of 20 nm. Within this range, the algorithm loops through every allowed pair of wavelengths and computes the difference in reflectance between them, $\Delta R = R(\lambda_{end}) - R(\lambda_{start})$. The maximum observed increase in reflectance for each spectrum ($\Delta R_{max}$) is recorded along with the $\lambda_{start}$ and $\lambda_{end}$ values. Spectra are then ranked in decreasing order of $\Delta R_{max}$. To obtain only step-function like spectra and exclude narrow absorption troughs, spectra were excluded if the reflectance at the midpoint of the interval ($\lambda_{start}$ to $\lambda_{end}$) was exceeded by any reflectance value for lower λ (including the region from 0–600 nm). Similarly, to avoid narrow reflectance peaks, spectra were excluded from our ranking if the reflectance at the midpoint of the interval was higher than any reflectance at greater λ (up to 1200 nm). This procedure excluded, for example, the rare-earth-element minerals monazite (a phosphate) and bastnaesite (a carbonate-fluoride), which show strong increases in reflectance at around 680 nm (with $\Delta R_{max}$ of 25.86 and 25.45 respectively) followed by even stronger decreases at around 730 nm. Finally, for further discussion in this paper we retain only those spectra with an $\Delta R_{max}$ of at least 14, i.e., those with the most pronounced edges.

Our final ranking was obtained from all minerals in the databases. However, only a subset of the USGS artificial materials are considered astrobiologically relevant: namely potassium ferrocyanide (included for its potential role in the origin of life (Toner & Catling, 2019; Sasselov et al., 2020)) and inorganic pigments, the latter including an iron oxide powder, Iron_Oxide SA-480665 gamma ASDHRa AREF (speclib07a rec 18816), regarded by Borges et al. (2024) as an environmentally appropriate false positive for the VRE/PRE. Minerals with the strongest positive slope are listed in **Table 1** and their reflectance spectra from 500 to 1200 are plotted in **Figure 1**. For comparison, the red edges of plants in the ECOSTRESS database with the top-ranked $\Delta R_{max}$ values are tabulated in **Supplementary Table 1** and visualized in **Supplementary Figure 1**.

*2.3 Simulating mineral red edge false positives on exoplanets with the NASA Planetary Spectrum Generator*

NASA's Planetary Spectrum Generator (PSG) is an online radiative transfer suite developed at NASA Goddard Space Flight Center, accessible at: https://psg.gsfc.nasa.gov/ (Villanueva et al., 2018). It is designed to simulate planetary spectra for a wide range of celestial bodies, including planets, exoplanets, comets, and other small bodies, across a broad spectrum from radio to ultraviolet wavelengths. PSG integrates advanced radiative transfer modeling with extensive spectroscopic databases and planetary ephemeris databases to perform detailed spectral simulations. High-resolution spectra are computed using line-by-line calculations, while moderate-resolution spectra use the correlated-k approach, which involves pre-computing opacities for a range of temperatures, pressures, and collisional regimes to provide a balance between computational efficiency and accuracy, especially for simulating large wavelength ranges where full line-by-line calculations are not practical (Villanueva et al., 2018). Beyond spectral generation, PSG facilitates the fitting of modeled spectra to observational data, enabling the determination of atmospheric species abundances, temperature and altitude profiles, and other key parameters (Villanueva et al., 2018).

We interacted with PSG primarily via remote instructions to the online server through the Application Program Interface (API). The simulation parameters are defined in a configuration file, encompassing atmospheric, object/geometry, and instrumental settings, as detailed in Villanueva et al. (2018). The simulation outputs include the generated spectrum (with units dependent on the simulation type), atmospheric component transmittances, and observational noise components within the selected wavelength range. The LUVOIR B-VIS template within the Extreme Coronagraph for Living Planetary Systems (ECLIPS) was used in PSG to generate simulated spectra because of its similarity to the proposed design of HWO. LUVOIR B, a concept for an 8-meter aperture space observatory, was designed to provide continuous spectral coverage from the far-ultraviolet to the near-infrared, with ECLIPS delivering data across three channels: UV, VIS, and NIR (LUVOIR Team, 2019). Within PSG, the VIS channel offers a resolution of up to 140 within the 515 nm to 1030 nm range and the NIR channel offers a resolution of up to 70 within the 1000 to 2000 nm range (Villanueva et al., 2018). The maximum coronagraph throughput of the planetary signature for LUVOIR B is noted as roughly 0.54, optimized to maximize planetary signature throughput while minimizing stellar flux (Villanueva et al., 2018). LUVOIR B's design incorporated ECLIPS to facilitate direct observations of Earth-like exoplanets, complemented by the option of an external starshade for enhanced contrast (LUVOIR Team, 2019).

Each simulated exoplanet is Earth-sized, with 1 g gravity, orbiting a G-type star at 1 AU. Simulations were conducted with and without an atmosphere, where the atmosphere is modeled after Earth's, with an identical level of oxygen, 1013 mbar surface pressure, and 60 atmospheric layers. Oxygen has a small effect on the magnitude of the recovered VRE/PRE because of the $O_2$ absorption features at ~690 and ~720 nm (see Discussion for consideration of the co-occurrence of oxygen and mineral false positives). The observation was simulated from 10 parsecs away at an observation time of an ideal 144,000 seconds (40 hours). The output radiance units are I/F. PSG derives its Earth-spectrum model from a combination of internal databases and external sources, primarily using the Modern-Era Retrospective Analysis for Research and Applications (MERRA-2) database for atmospheric profiles and molecular abundances. MERRA-2 integrates meteorological data from numerous sources (Villanueva et al., 2018).

The "Abiotic Earth" surface used in some simulations was defined as 71% seawater (*Open_ocean_sw2_27262_USGS[0.21-2.98um]),* 14.28% granular sediment (*SM-CMP-031-C1SM31_RELAB[0.60-2.70um Sandstone Limestone Soil STN 17 B Sediment]),* 10.39% granite to represent mountain/bare rock (*RA-REA-015-C1RA15_RELAB[0.35-2.60um Granite Gneiss E4 144 Rock]), and* 4.33% water ice to represent ice/snow *(AFCRL_Ice_HRI[0.20-30000.00um]).*

## 3. Results

### *3.1 Identification of VRE/PRE-like features in minerals*

Our algorithmic search through the ECOSTRESS and USGS spectroscopic databases identified edge-like increases in reflectance between 600 and 800 nm in numerous minerals. Minerals with the strongest positive slope in this region are listed in **Table 1** and their reflectance spectra from 500 to 1200 are plotted in **Figure 1** on absolute scale (left) as well as on a normalized scale (right) to show the similarities of the slope characteristics clearly.

Note that many of the minerals with the strongest edge-like increases in reflectance in the relevant region are sulfides, including sulfides of lead, copper, mercury, arsenic and zinc. Among these are the edge-like profiles of cinnabar (HgS) and elemental sulfur (S) (see also Seager et al., 2005). The blue feldspathoid tectosilicate sodalite is also prominent, as is lazurite, which is a sulfur-rich variety of sodalite. The weakest edge-like profiles retained by our ranking were those of the nesosilicate zircon (a ubiquitous but very minor component of granitic rocks), the phyllosilicate nontronite (an iron-rich swelling clay common on Earth and Mars) and quartz, which is one of the most abundant minerals on Earth's surface.

**Table 1**

| Mineral (F = fine; M = medium; C = coarse) | Class | Database | $\Delta R_{max}$ | Interval width, nm (λend–λstart) | Interval midpoint, nm |
|---|---|---|---|---|---|
| Stibnite, F | Sulfide | ECOSTRESS | 48.71 | 80 | 760 |
| Chalcopyrite, M | Sulfide | ECOSTRESS | 29.97 | 80 | 760 |
| Lazurite, M | Tectosilicate | USGS | 26.50 | 80 | 695 |
| Cinnabar, C | Sulfide | USGS | 25.74 | 80 | 640 |
| Sodalite, C | Tectosilicate | ECOSTRESS | 25.68 | 80 | 695 |
| Realgar, C | Sulfide | ECOSTRESS | 21.93 | 80 | 640 |
| Zircon, C | Nesosilicate | ECOSTRESS | 18.91 | 80 | 692 |
| Nontronite, F | Phyllosilicate | ECOSTRESS | 16.69 | 80 | 708 |
| Quartz, C | Tectosilicate | USGS | 14.83 | 79 | 753 |

***Table 1.*** *Minerals identified as possessing edge-like reflectance features in the 600–800 nm region of the spectrum. To avoid repetition, those minerals for which multiple sources of data were available (e.g., different grain sizes or instruments) are represented in this table only by the instance with the strongest slope. The chemical composition of each mineral is as follows, all sourced from Mindat.org [https://www.mindat.org/]: stibnite =* $Sb_2S_3$*; chalcopyrite =* $CuFeS_2$*; lazurite =* $Na_7Ca(Al_6Si_6O_{24})(SO_4)(S_3)\cdot H2O$*; cinnabar = HgS; sodalite =* $Na_4(Si_3Al_3)O_{12}Cl$*; realgar =* $As_4S_4$*; zircon =* $Zr(SiO_4)$*; nontronite =* $Na_{0.3}Fe_2((Si,Al)_4O_{10})(OH)_2\cdot nH_2O$*; quartz =* $SiO_2$

*3.2 Edge-like feature in cyanide salts and artificial pigments*

Our search through the USGS artificial materials database revealed that spectra of many cyanide compounds/salts, and artificial inorganic pigments have features resembling the VRE/PRE. The compounds with the strongest positive slope between 600 and 800 nm are listed in **Table 2.** The reflectance spectra between 500 to 1200 of selected artificial materials are illustrated in **Figure 1.**

**Table 2**

| Material | Compound | ΔRmax | Interval, nm (λend–λstart | Interval midpoint |
|---|---|---|---|---|
| Cadmium red paint | Cd(S,Se) | 56.29 | 80 | 641 |
| Potassium ferrocyanide | $K_4Fe(CN)_6$ | 36.33 | 80 | 640 |
| Cadmium orange paint | CdS | 26.73 | 80 | 640 |
| Vermillion paint | HgS | 20.44 | 80 | 749 |
| Lazurite paint | $(Na,Ca)_8Al_6Si_6O_24(SO_4,S)$ | 19.67 | 80 | 640 |
| Iron (III) Oxide | $Fe_2O_3$ | 15.52 | 80 | 687 |
| Ultramarine paint | $Na_7Al_6Si_6O_{24}S_3$ | 14.66 | 76 | 730 |

***Table 2.*** *Artificial materials including cyanide compounds and inorganic pigments (paints) identified as possessing edge-like reflectance features in the 600–800 nm region of the spectrum. All spectra are sourced from the USGS database. The chemical compositions of paints are sourced from Colour Lex [https://colourlex.com/project/cobalt-blue/].*

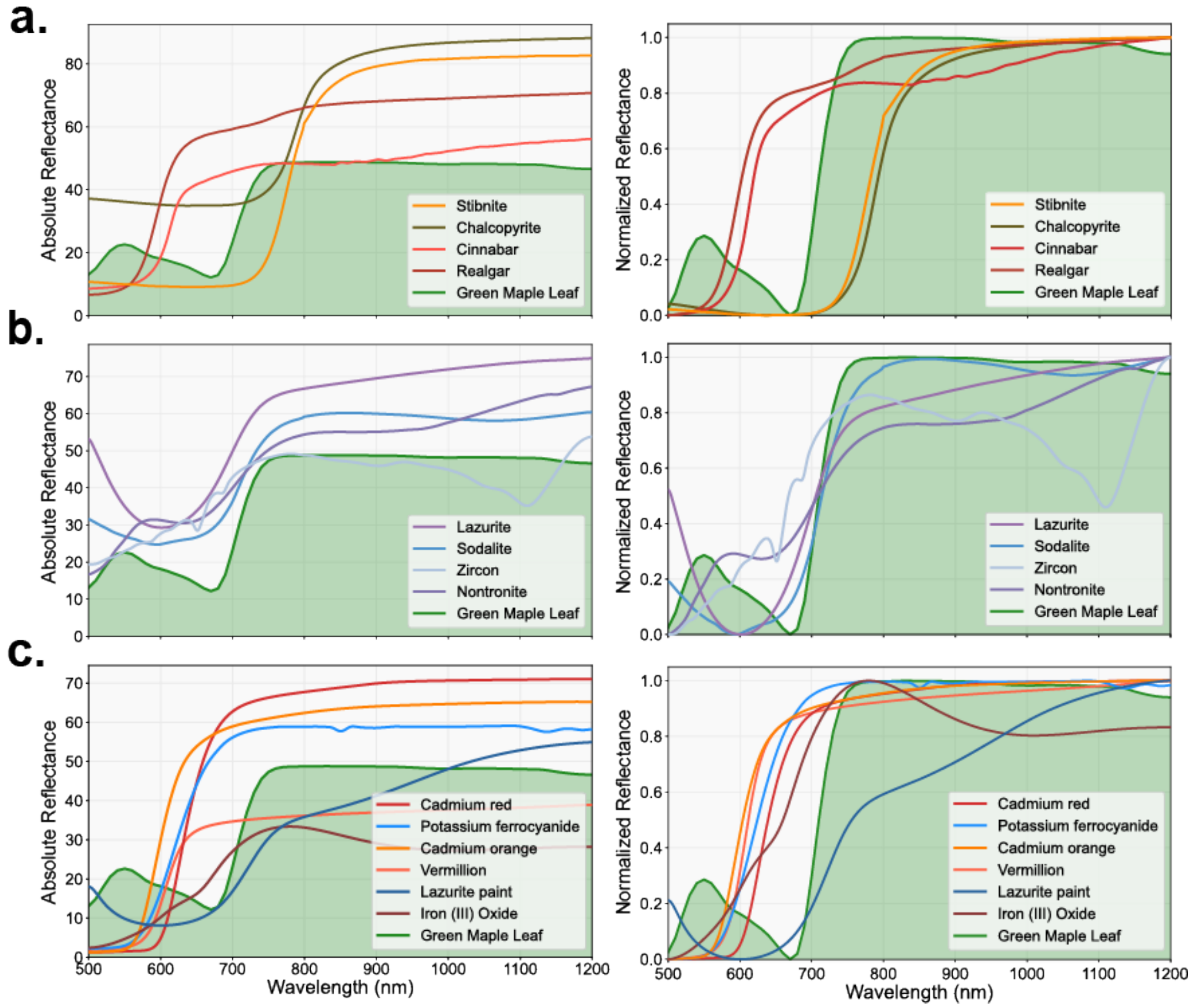

***Figure 1. Comparison of (left) absolute and (right) normalized mineral and artificial reflectance spectra with the photosynthetic red edge.*** *Spectra on the right were vector-normalized here to facilitate visualization of spectral features rather than overall reflectance. All spectra were also subjected to Gaussian smoothing with standard deviation 3. (a) sulfide mineral spectra from* ***Table 1*** *(database source: stibnite, ECOSTRESS | chalcopyrite, ECOSTRESS | cinnabar, USGS | realgar, ECOSTRESS) alongside the green leaf spectrum used to model the VRE/PRE; (b) silicate mineral spectra from* ***Table 1*** *(database source: lazurite, USGS | sodalite, ECOSTRESS | zircon, ECOSTRESS | nontronite, ECOSTRESS)*

*with the VRE/PRE; and (c) artificial material spectra from* ***Table 2*** *(database source: all USGS) with the VRE/PRE.*

*3.3 Detectability of false positives in telescope simulations*

How much of a planet's surface would need to be covered by a given mineral to match the red edge slope of a given amount of photosynthetic biomass? To answer this question, we used the LUVOIR B telescope instrument model within the NASA Planetary Spectrum Generator to simulate observations of an Earth twin at a distance of 10 parsecs (see Methods for details), with modified surface compositions. We modeled spectra at a spectral resolution of 140 of the simulated planet with varying amounts of photosynthetic biomass, represented for simplicity by the PSG "Green Maple Leaf" spectrum. Although this is not a realistic spectral substitute for all life on Earth's surface (see e.g., O'Malley-James & Kaltenegger 2017 for changes of VRE/PRE over geological time), it's very strong VRE/PRE leads to conservative estimates of the potential false-positive effect of minerals. In our simulations, we covered 0 to 100 % of the planet's surface with this material (at 10% increments) and the remainder with "Abiotic Earth" (a mixture of water, sediment, rock, and snow/ice; see Methods). We then repeated this process using the strongest mineral false positives listed in **Table 1** instead of photosynthetic biomass, and compared the biomass and mineral-dominated planets. **Figure 2** shows the results of these comparisons, including for example a 100% "Abiotic Earth" planet compared to a uniform surface with Earth's albedo, and also highlights the similarity in the VRE/PRE region of the spectrum between a planet with 22% maple leaf surface coverage and a planet with 70% lazurite surface coverage. These simulations included an Earth-like atmosphere; see **Supplementary Figure 2** for an analogous comparison of "Abiotic Earth", 22% maple leaf and 70% lazurite with no atmosphere.

**Figures 1** and **2** show that several minerals with VRE/PRE-like edges also display a greater overall increase in reflectance between visible and near-infrared light than vegetation, even though the slope of the increase is lower. These minerals would represent especially potent false positives for low-resolution observations in which the slope near 700 nm cannot be resolved. Nevertheless, when this region of the spectrum is well resolved, the observable red edges in the simulated spectra display smaller $\Delta R_{max}$ values (between 685 and 715 nm) for Earth-twins covered in the mineral than for Earth-twins covered in the green maple leaf material. This is illustrated by **Figure 3**: in both no-atmosphere (top) and Earth-like atmosphere simulations (bottom), minerals generally exhibit shallower slopes at ~700 nm than photosynthetic biomass. Thus, they would need to occupy a larger portion of the surface than biomass to replicate an equivalent VRE/PRE signal. The identification of any pigments is easier without an atmosphere, as anticipated.

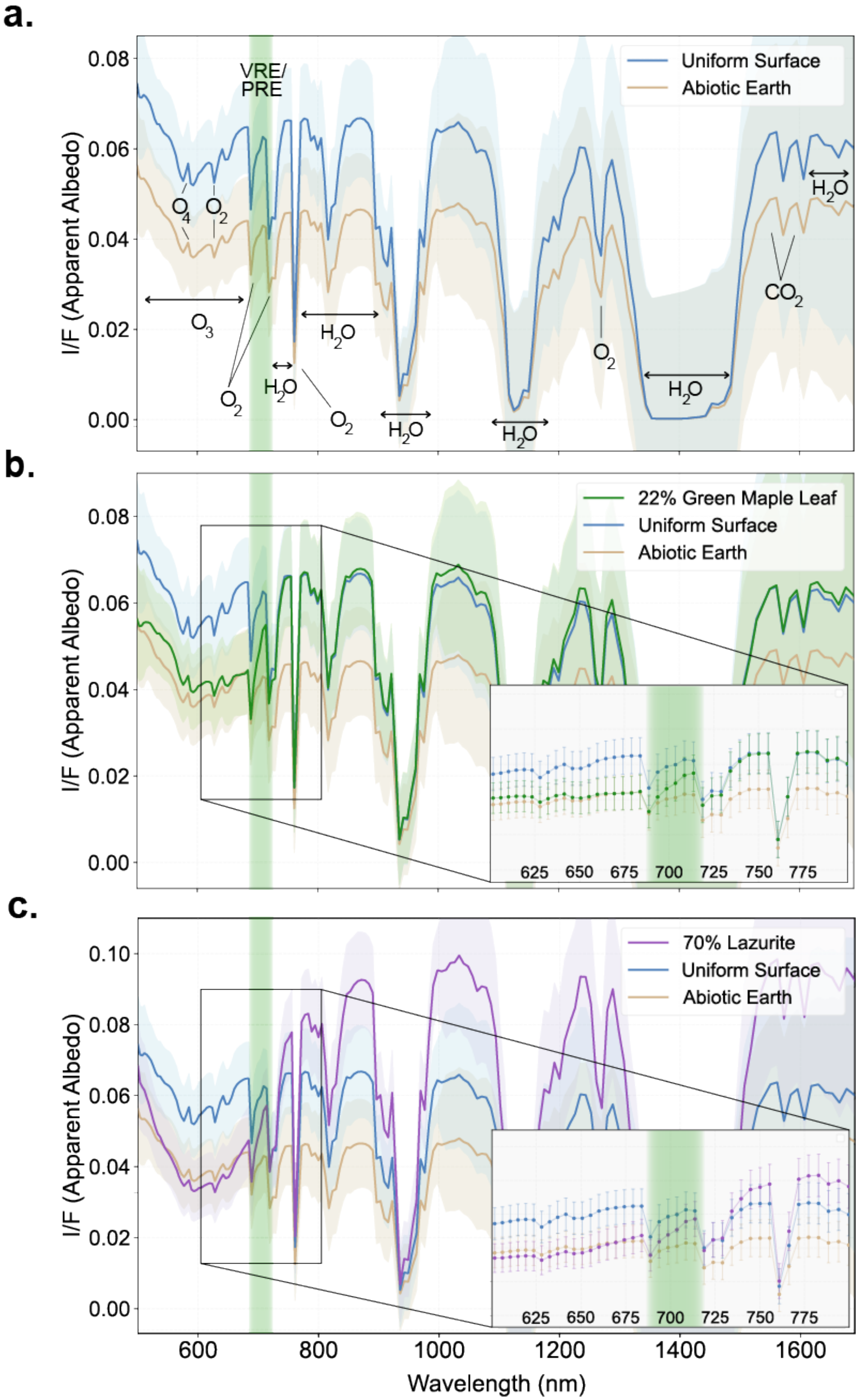

a.
Uniform Surface
Abiotic Earth
VRE/
PRE
I/F (Apparent Albedo)
0.08
0.06
0.04
0.02
0.00
O4
O2
O3
O2
H2O
O2
H2O
H2O
H2O
O2
H2O
CO2
H2O
b.
22% Green Maple Leaf
Uniform Surface
Abiotic Earth
625
650
675
700
725
750
775
c.
70% Lazurite
Uniform Surface
Abiotic Earth
0.10
600
800
1000
1200
1400
1600
Wavelength (nm)

***Figure 2.*** *Simulated reflectance spectra (I/F) of alternative Earths. The shaded errors represent telescope noise in an observation simulated with PSG with 8 hours of exposure time (LUVOIR B-VIS instrument). All simulations use modern Earth's atmosphere with 21% oxygen. (a) Earth with neither vegetation nor VRE/PRE-like mineral edges. The "uniform surface" spectrum was generated using a featureless surface (uniform albedo 0.308), to highlight the atmospheric features. The "Abiotic Earth" spectrum includes oceans, ice, deserts and granitic rock in Earth-like proportions. (b) Earth with a surface comprising 22% green maple leaf, with the remainder being "Abiotic Earth". (c) Earth with a surface comprising 70% lazurite, with the remainder being "Abiotic Earth". Atmospheric components responsible for key spectral features are labeled in each panel, sourced from Claudi and Alei (2019). Note the* $O_2$ *absorption features at ~690 and ~720 nm, bracketing the VRE/PRE. These surfaces are modeled without atmospheres in* ***Supplementary Figure 2****.*

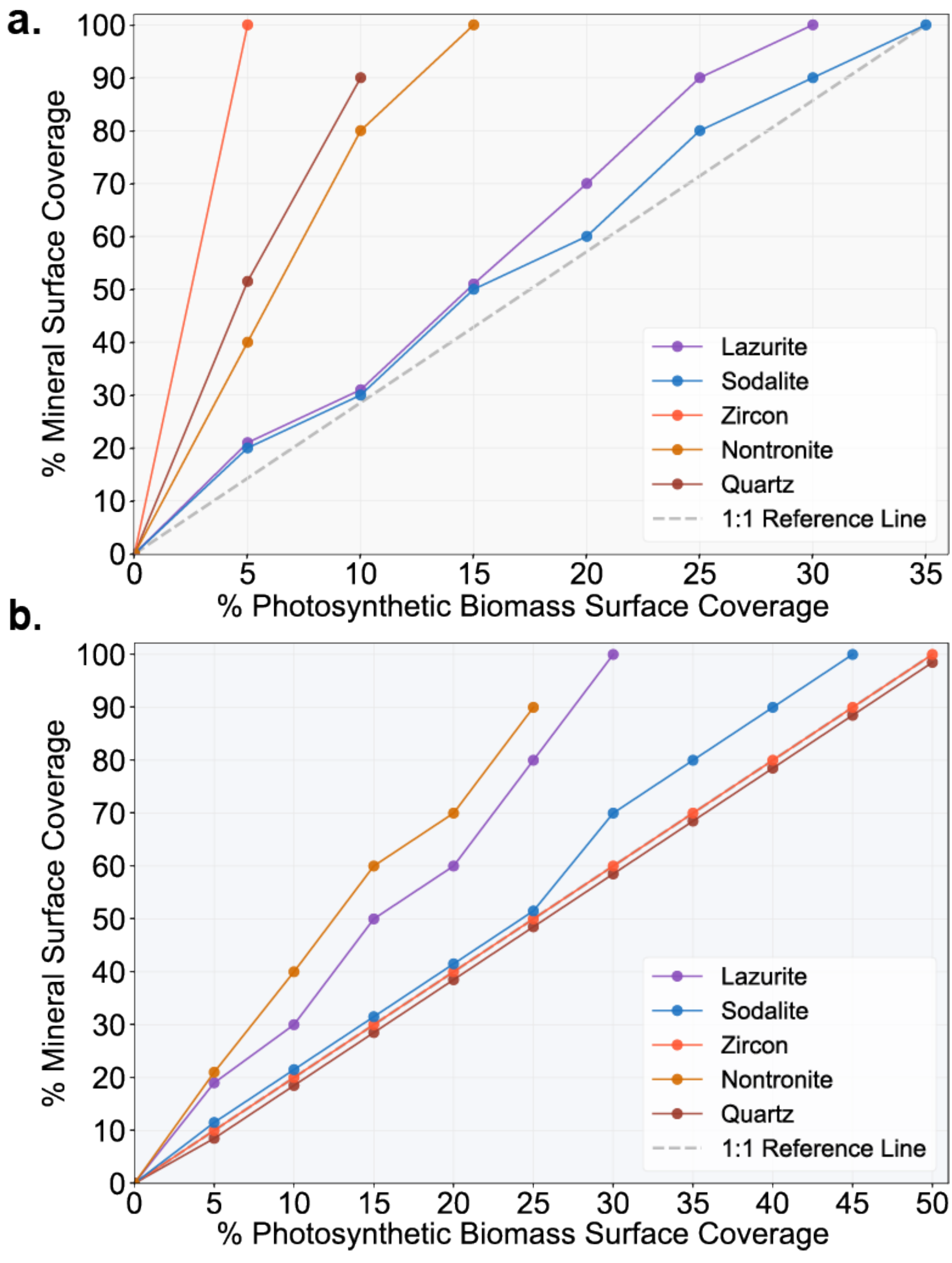

***Figure 3.*** *Minerals as a false positive for the VRE/PRE in simulated exoplanet spectra without instrument noise from PSG for* ***(a)*** *No-atmosphere cases* ***(b)*** *Earth-like atmosphere cases. Surface coverage (% planet surface) by minerals required to match the apparent red edge slope ($\Delta R_{max}$ in the 685–715 nm window) of photosynthetic biomass, for different amounts of*

*biomass surface coverage. The remaining surface was identical to "Abiotic Earth" shown in **Figure 2**. Data were obtained using the simulated LUVOIR B-VIS instrument in the PSG for an Earth-like planet orbiting a Sun-like star at 1.A.U, viewed from a distance of 10 parsecs. Sulfide minerals were excluded because their "edges" are outside of the 685–715 nm window (as shown by **Figure 1**). The green maple leaf spectrum was used as the example for photosynthetic biomass.*

## 4.0 Discussion

Our results suggest that more minerals than previously recognized can mimic VRE/PRE-like features, expanding the range of potential false positives associated with future direct imaging of Earth-like exoplanets. Previously, Seager et al. (2005) identified elemental sulfur and cinnabar as potential false positives and Borges et al. (2024) added iron oxide. We performed an in-depth search for natural and artificial materials and identified additional minerals that could be mistaken for biopigments (see **Table 1**): We add several other sulfides (realgar, stibnite, chalcopyrite) and silicates (especially lazurite and sodalite) to the list and note that zircon, nontronite and even quartz also pose a minor risk of false positives. Lazurite and sodalite match the spectral shape of the VRE/PRE especially well (**Figure 1**) although the magnitude of the edge is smaller. Lazurite and especially sodalite can effectively mimic the VRE/PRE in simulated telescope data (**Figure 3**).

### *4.1 Surface composition and mineral assemblages*

While this study has compared the spectroscopic signals of different materials with coverage ranging from 0 to 100% of a planet's surface, real surface coverage is unlikely to approach 100% either for biomass or for a single mineral. Indeed, no single mineral (other than ices or salts) is likely to occur over a high proportion of the surface of a planetary body except as one component in a milieu with other minerals (i.e., a mineral assemblage). In igneous rocks, particular minerals typically co-occur with others as part of an assemblage, where the total assemblage is, to a first approximation, the predictable consequence of thermodynamic equilibration between solid and molten rock. Fractional crystallization can produce mineral assemblages out of equilibrium with the original melt, as can metamorphic alteration and the vagaries of erosion and redeposition leading to sedimentary rocks. In some conditions these disequilibrium processes can generate mineral assemblages dominated by one or a few components such as the basalt making up the lunar highlands, which is more than 90% composed of a single mineral (the Ca-plagioclase feldspar, anorthite) (Michaut & Neufeld, 2022). Nevertheless, even in metamorphic and igneous rocks, most minerals are found most commonly in association with only a few other minerals. In sandstones, quartz is commonly paired with feldspar and clay; in basalts, plagioclase feldspar is usually paired with olivine, and so on. Although our approach in this paper has the virtue of simplicity, future work should seek to identify spectral reflectance features of realistic mineral assemblages in which false-positive minerals may occur; such features may help to disambiguate false positives from biosignatures. Equally, the use of realistic proxies for green biomass (rather than maple leaves) would enhance the realism of future simulations.

### *4.2. Choice of spectral features of interest*

Our analysis focused on the visible and NIR range, avoiding the increased telescope noise in the PSG LUVOIR-B instrument model above 1000 nm. Because of the rules we used to search the spectral databases, the minerals we highlight as false positives do not show any sharp features (other than the VRE/PRE-like edges) in the 500 nm to 1200 nm range. However, they may show features in other parts of the spectrum that, if resolvable above instrument noise, could help to disambiguate them from biomass.

Another limitation of the present work is that we have only considered the canonical VRE/PRE at 700 nm. On other planets, photosynthesizers may also evolve different absorption and reflection behavior depending on the spectrum of sunlight they receive, itself a function of stellar type and age (e.g., Kiang et al., 2007; O'Malley-James et al., 2012). Even on Earth, the spectroscopic fingerprints of real biomass are diverse. For instance, Hegde et al. (2015) show that many pigmented microorganisms have other edge-like spectral features that should be considered in the interpretation of future exoplanet reflectance data (Hegde et al., 2015). Coelho et al. (2024) highlight that early Earth may have been dominated by purple sulfur bacteria and purple non-sulfur bacteria with purple carotenoid and bacteriochlorophyll-based pigments, which absorb and reflect light in the 750 to 1100 nm range where M-stars radiate most intensely, potentially allowing such organisms to thrive on planets orbiting these stars (Coelho et al., 2024). There is also the possibility of biofluorescence (O'Malley-James & Kaltenegger, 2017). Thus, future work must consider how minerals may represent false positives for reflectance biosignatures other than the VRE/PRE. Our approach can readily be adapted to search for edge-like features in mineral spectra outside the 600–800 nm range prioritized in this study.

The present study did not consider the dynamic nature of planetary spectra. Earth's observable red edge varies according to the proportion of land vs. ocean visible to the observer at the time of observation, as well as to seasonality and the extent of ice and cloud coverage (e.g., Hamdani et al., 2006; Montañés-Rodriguez et al., 2006; Arnold, 2008). However, some of the same temporal variability might also be expected for mineral-rich regions that rotate in and out of view and are seasonally covered by ice, cloud, or dust (as on Mars).

The photosynthetic red edge observed on modern Earth is accompanied by the presence of atmospheric oxygen. Co-occurrence of a VRE/PRE and oxygen absorption features ($O_2$ or $O_3$) would be more compelling evidence of life than either of these alone. However, a VRE/PRE could exist without a co-occurring oxygen signature, as photosynthetic life evolved long before significant atmospheric oxygen accumulated, which required enhanced burial of organic carbon and the removal of other sinks for oxygen (e.g., Lyons et al., 2014; Deitrick & Goldblatt, 2023). Even following the Great Oxygenation Event (GOE) ~2.4 billion years ago, oxygen and ozone levels remained between one and ten percent of modern levels for nearly two billion years and may have been too low to be detected remotely even when green biomass was abundant (e.g., Reinhard et al., 2017; Mills et al., 2023). Although it is possible that ozone would have been detectable without spectrally apparent oxygen during the mid-Proterozoic, it was certainly detectable by the end of the eon (e.g., Kaltenegger et al., 2007, 2009; Reinhard et al., 2017). Numerous studies have also demonstrated scenarios where oxygen could be present without a prominent VRE/PRE (Cockell et al., 2009; Schwieterman et al., 2018). Therefore, oxygen and the VRE/PRE are not always coupled; the presence or absence of oxygen cannot be reliably used as a sole disambiguating factor to distinguish between a true VRE/PRE and mineralogical false positives.

The occurrence of a reflectance edge at ~650 nm in the spectrum of potassium ferrocyanide is also notable given that surface deposits of ferrocyanide salts have been predicted to occur on Earth-like planets as a consequence of prebiotic ('cyanosulfidic') chemistry (e.g., Toner & Catling, 2019; Sasselov et al., 2020). This chemistry is argued to have been driven both by endogenous organic synthesis and by exogenous delivery of hydrogen cyanide in cometary impactors, after which ferrocyanide salts would accumulate in alkaline lakes (e.g., Toner & Catling, 2019). More broadly, cyanide compounds have long been considered important feedstocks for the origin of life (for example, the famous Strecker synthesis of amino acids proceeds via the reaction of an aldehyde with hydrogen cyanide) (e.g., Ishitani et al., 2000). It appears to be a coincidence that a canonical signature of life, the VRE/PRE, may be mimicked spectrally by a prebiotically important compound. However, the potentially misleading capacity for life's precursors to resemble life has been noted in other contexts (e.g., McMahon & Jordan, 2022).

## 5. Conclusion

Biopigments on exoplanet surfaces are a critical target in the search for life in the cosmos. However, no in-depth study has previously been undertaken to identify minerals that could mimic the "vegetation" or "photosynthesis red edge" (VRE or PRE), a potential biosignature associated with photosynthetic biopigments. This study has filled this gap and demonstrated that several minerals and artificial materials exhibit VRE/PRE-like spectral features, specifically sharp reflectance edges in the 600–800 nm range. Through algorithmic analysis of extensive spectral databases and simulations for the anticipated HWO mission, we have identified a range of sulfides and silicates, as well as the prebiotically important compound potassium ferrocyanide, that present significant false-positive risks for future exoplanet biosignature detection.

Several minerals with VRE/PRE-like edges also show a greater overall increase in reflectance between visible and near-infrared light than vegetation. These minerals represent especially potent false positives in low-resolution or high-noise observations where the slope near 700 nm cannot be clearly seen. Where this slope can be resolved, however, (as in our simulated telescope observations), we find that minerals able to mimic the VRE/PRE need to cover a larger area of a planet to produce a similarly strong slope at ~700 nm. The probability may be low that large proportions of real planetary surfaces would be dominated by any of these materials. However, the same may be said of biological materials, and the possibility of confusion increases for lower biopigment coverage.

It is now critical to identify mixtures of minerals and other natural substances that could mimic a wide variety of surface biopigments, and to develop new strategies for disambiguating abiotic and biotic materials in reflectance spectra. Such strategies are especially needed for evaluating candidate reflectance biosignatures that occur without strong corroborating evidence for the presence of life (such as atmospheric biosignatures). Further work on false positives in exoplanet astrobiology will not undermine the search for life in the cosmos but, on the contrary, place it on a surer footing.

## Acknowledgements

This work was supported by a NERC E4 doctoral studentship awarded to M. B. Parkinson, and an Edinburgh–Cornell Strategic Collaboration Award to L. Kaltenegger, S. McMahon, and B. Biller. The authors thank C. Loron for his assistance with Quasar, R. Ramsay for his expertise in spectroscopic measurements, and G. Villanueva and V. Kofman for their help in running PSG simulations.